\documentclass{article}

\usepackage[aitw]{iclr2026_conference}
\usepackage{times}
\usepackage{amssymb}
\usepackage{amsmath}
\usepackage{booktabs}
\usepackage{tabularx}
\usepackage{graphicx}
\usepackage{float}
\usepackage{tikz}
\usetikzlibrary{positioning,arrows.meta,shapes,calc}
\usepackage{enumitem}
\usepackage{url}
\usepackage{hyperref}
\usepackage[capitalize]{cleveref}
\crefname{section}{Appendix}{Appendices}
\crefname{section}{Appendix}{Appendices}
\usepackage{algorithm}
\usepackage{algpseudocode}
\usepackage{enumerate}
\usepackage{comment}
\usepackage{listings}
\usepackage{nicefrac}
\usepackage[T1]{fontenc}
\usepackage{mdframed}
\usepackage{verbatim}
\usepackage{fvextra}
\newcommand{\displayprompt}[1]{%
  \begin{mdframed}[backgroundcolor=gray!2, linecolor=black]
    \VerbatimInput[
      breaklines=true,
      breakanywhere=true,
      fontsize=\tiny,
      baselinestretch=0.5,
      breaksymbolleft=
    ]{#1}
  \end{mdframed}
}

\title{Can AI Agents Agree?}

\author{Frédéric Berdoz, Leonardo Rugli, Roger Wattenhofer \\
ETH Zurich, Switzerland\\
\{fberdoz, lrugli, wattenhofer\}@ethz.ch \\
}

\iclrfinalcopy
\begin{document}

\maketitle

\begin{abstract}
Large language models are increasingly deployed as cooperating agents, yet their behavior in adversarial consensus settings has not been systematically studied. We evaluate LLM-based agents on a Byzantine consensus game over scalar values using a synchronous all-to-all simulation. We test consensus in a no-stake setting where agents have no preferences over the final value, so evaluation focuses on agreement rather than value optimality. Across hundreds of simulations spanning model sizes, group sizes, and Byzantine fractions, we find that valid agreement is not reliable even in benign settings and degrades as group size grows. Introducing a small number of Byzantine agents further reduces success. Failures are dominated by loss of liveness, such as timeouts and stalled convergence, rather than subtle value corruption. Overall, the results suggest that reliable agreement is not yet a dependable emergent capability of current LLM-agent groups even in no-stake settings, raising caution for deployments that rely on robust coordination.
\end{abstract}

\begin{figure}[ht]
\centering
\begin{tikzpicture}[
    font=\small,
    node distance=1cm,
    >={Stealth[length=2mm]},
    every node/.style={font=\small}
]

\def\ringRadius{2.2}
\coordinate (ringCenter) at (0, 0);

\node[circle, fill=blue!30, draw=blue!70, very thick, minimum size=14mm, inner sep=0pt]
    (Ai) at ($(ringCenter)+(0:\ringRadius)$) {\footnotesize Agent $i$};

\node[circle, fill=blue!20, draw=blue!60, minimum size=10mm, inner sep=0pt]
    (A1) at ($(ringCenter)+(90:\ringRadius)$) {};
\node[circle, fill=blue!20, draw=blue!60, minimum size=10mm, inner sep=0pt]
    (A3) at ($(ringCenter)+(180:\ringRadius)$) {};
\node[circle, fill=blue!20, draw=blue!60, minimum size=10mm, inner sep=0pt]
    (A4) at ($(ringCenter)+(225:\ringRadius)$) {};
\node[circle, fill=blue!20, draw=blue!60, minimum size=10mm, inner sep=0pt]
    (A5) at ($(ringCenter)+(270:\ringRadius)$) {};
\node[circle, fill=blue!20, draw=blue!60, minimum size=10mm, inner sep=0pt]
    (A6) at ($(ringCenter)+(315:\ringRadius)$) {};

\node[circle, fill=red!20, draw=red!60, dashed, thick, minimum size=10mm, inner sep=0pt]
    (B7) at ($(ringCenter)+(45:\ringRadius)$) {};
\node[circle, fill=red!20, draw=red!60, dashed, thick, minimum size=10mm, inner sep=0pt]
    (B2) at ($(ringCenter)+(135:\ringRadius)$) {};

\draw[gray!40, thin] (A1) -- (B2);
\draw[gray!40, thin] (B2) -- (A3);
\draw[gray!40, thin] (A3) -- (A4);
\draw[gray!40, thin] (A4) -- (A5);
\draw[gray!40, thin] (A5) -- (A6);
\draw[gray!40, thin] (A6) -- (Ai);
\draw[gray!40, thin] (Ai) -- (B7);
\draw[gray!40, thin] (B7) -- (A1);

\draw[->, thick, blue!70] (Ai.west) to[bend left=18]
    node[pos=0.45, below, font=\footnotesize, sloped] {1. Broadcast} (A4.east);

\draw[->, thick, blue!70] (A4.north east) to[bend left=18]
    node[midway, above, font=\footnotesize, sloped] {2. Receive} (Ai.north west);

\draw[->, thick, blue!70] (Ai.west) to[bend right=35]
    node[midway, below, font=\footnotesize, sloped] {3. Vote} (A4.east);

\draw[->, thick, blue!70]
  (Ai.110) to[out=120, in=40, looseness=10]
  node[pos=0.7, above=9pt, font=\footnotesize]
       {4. Private strategy}
  (Ai.70);

\draw[->, thin, blue!40] (A1) to[bend right=20] (A6);
\draw[->, thin, red!40] (B2) to[bend left=15] (A4);
\draw[->, thin, red!40] (B7) to[bend left=15] (A3);

\node[draw=gray!70, rounded corners=4pt, fill=gray!5,
      minimum width=3.5cm, minimum height=2.9cm]
      (stateBox) at (5.5, 0.5) {};

\node[anchor=north, font=\small\bfseries] at (stateBox.north) {Internal State};

\node[anchor=north west, font=\footnotesize, text width=2.9cm, align=left]
    at ($(stateBox.north west)+(0.2,-0.5)$) {
    $v_i^{(t)}$: proposal\\[4pt]
    $h_i^{(t)}$: history\\[4pt]
    $\texttt{stop}_i$: vote\\[4pt]
    $r_i^{(t)}$: private strategy
};

\draw[<->, gray!50, thick, dashed] (Ai.east) -- (stateBox.west);

\node[circle, fill=blue!20, draw=blue!60, minimum size=6mm, inner sep=0pt]
    (legH) at (4.0, -2.0) {};
\node[right=0.3cm of legH, font=\scriptsize] {Honest};

\node[circle, fill=red!20, draw=red!60, dashed, minimum size=6mm, inner sep=0pt]
    (legB) at (6.0, -2.0) {};
\node[right=0.3cm of legB, font=\scriptsize] {Byzantine};

\end{tikzpicture}
\caption{Byzantine consensus game with honest and byzantine LLM agents on a synchronous all-to-all network in one round of interaction. The highlighted agent $i$ broadcasts a scalar proposal and justification, receives messages from peers, and emits a termination decision $\{\texttt{vote}, \texttt{continue}\}$. For clarity, only a subset of message arrows is shown.}
\label{fig:bcg_overview}
\end{figure}

\section{Introduction}

Large language models are increasingly deployed as autonomous agents that collaborate on planning, coding, reasoning, and many other tasks. In some applications, a group of agents must reach a common decision rather than producing independent answers. When some agents malfunction or behave strategically, the ability of the group to still reach agreement becomes an explicit design requirement rather than a convenience.
Classical Byzantine fault-tolerant consensus provides strong guarantees for deterministic algorithms~\citep{pease1980reaching, lamport1982byzantine}, but it is unclear how these guarantees translate when decisions are produced by stochastic, prompt-driven LLMs. We therefore study a simple \emph{no-stake} scalar consensus task in which a group of agents repeatedly propose values and exchange messages over a synchronous all-to-all network, and in which agents do not have specific preferences over the final value. A subset of agents may behave in a Byzantine fashion, arbitrarily attempting to disrupt or bias consensus. Each agent is implemented as an LLM-driven policy that receives a compact textual history and outputs a proposal and justification. \cref{fig:bcg_overview} illustrates the game structure.

We summarize our contributions as follows:
\begin{itemize}[nosep,leftmargin=*]
    \item A capability study of LLM-agent consensus across two models and group sizes in benign settings.
    \item A robustness study showing that even a single Byzantine agent collapses consensus success.
    \item An analysis revealing that failures are dominated by liveness loss, not value corruption.
\end{itemize}

\section{Related Work}

LLMs are increasingly deployed as interacting agents~\citep{li2023camel,wu2024autogen,liu2024agentbench}, and recent work documents common multi-agent failure patterns~\citep{cemri2025why}. Several approaches target robustness to malicious agents via coordination mechanisms~\citep{chen2024blockagents,jo2025byzantinerobust}. Closest to our setting, \citet{chen2023multiagent} study benign numeric consensus via LLM negotiation and analyze the effects of agent number, personality, and topology. Our focus differs in three ways. First, we introduce a controlled Byzantine fraction. Second, we enforce a validity constraint requiring the decided value to be an initial honest proposal. Third, we separate validity from liveness and show that failures are dominated by liveness loss. Complementary work frames multi-agent reliability through a Byzantine fault-tolerance lens and proposes aggregation mechanisms such as confidence-weighted consensus~\citep{zheng2025rethinking}. Concurrently, \citet{grotschla2025agentsnet} also observe that multi-agent LLM coordination degrades as network size increases. In a Mixture-of-Agents setting, \citet{wolf2025this} similarly find that a single deceptive agent suffices to nullify performance gains.

\begin{figure}[t]
    \centering
    \includegraphics[scale=0.5]{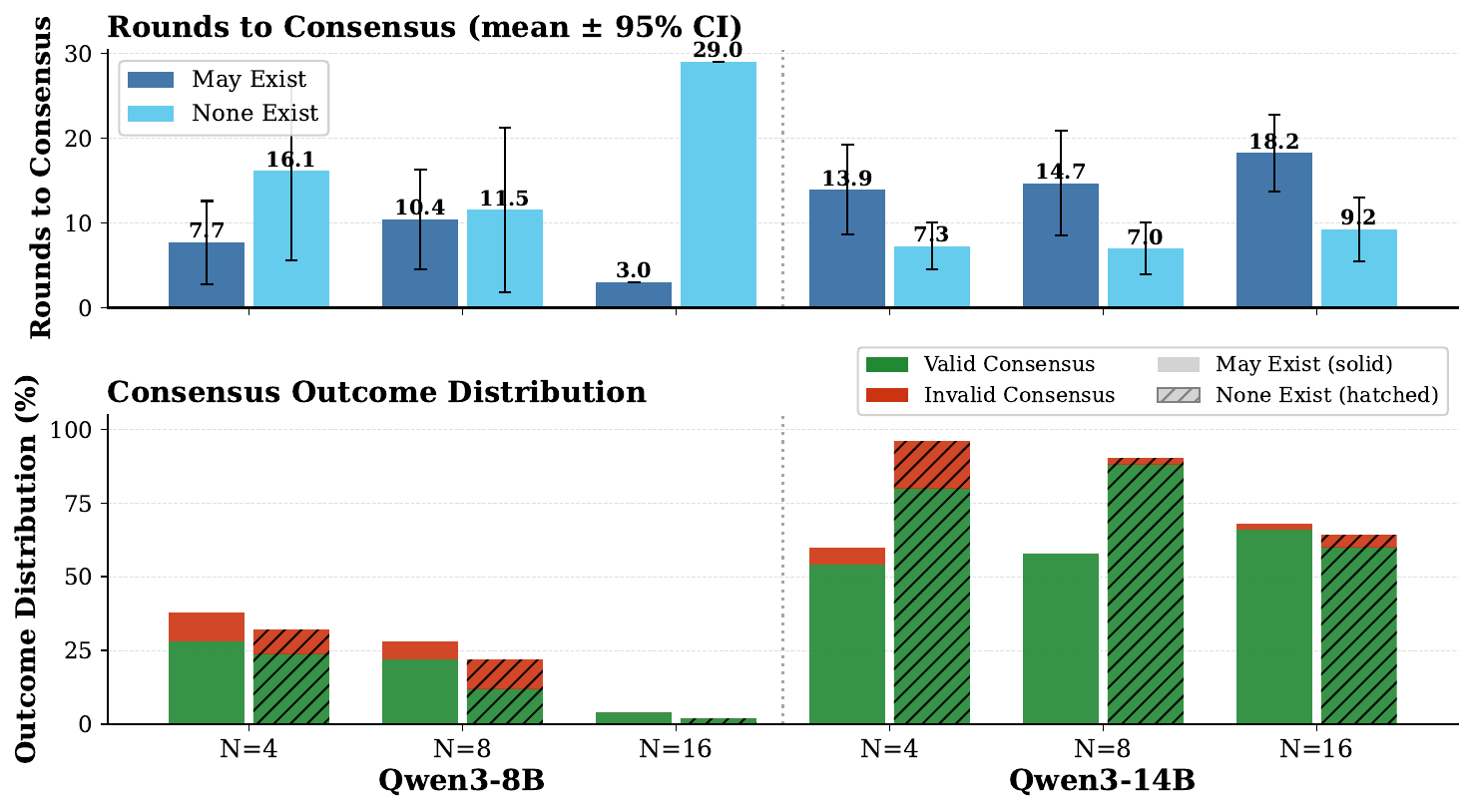}
    \caption{Consensus performance without Byzantine agents \(B=0\) for Qwen3-8B and Qwen3-14B across $N \in \{4,8,16\}$ and two prompt variants (with vs.\ without mentioning possible Byzantine agents). Error bars show 95\% Wilson confidence intervals. Qwen3-14B reaches valid consensus more often, adversary-free prompts improve liveness, and larger groups slow and weaken consensus.}
    \label{fig:combined_rounds_outcomes}
\end{figure}

\begin{figure}[t]
  \centering
  \includegraphics[width=1\linewidth]{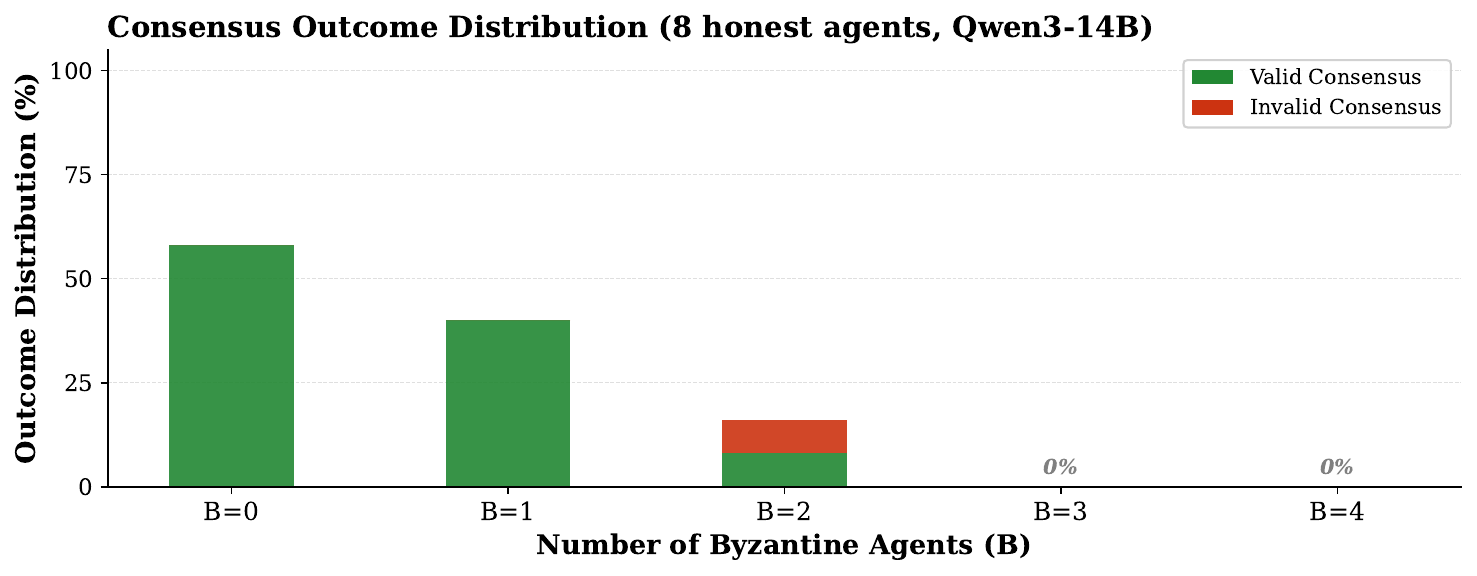}
  \caption{Effect of Byzantine agents on Qwen3-14B consensus with eight honest agents. Bars show the distribution of valid vs.\ invalid consensus outcomes over 25 runs per configuration; missing bars correspond to 0\% consensus (all timeouts).}
  \label{fig:q2_byzantine_agents_comparison}
\end{figure}

\section{Method}
\label{sec:method}

\paragraph{Setting.}
We consider \(N\) agents communicating over a synchronous all-to-all network for rounds \(t=1,\ldots,T_{\max}\).
A fraction \(f \in [0,\nicefrac{1}{3}]\) of agents are Byzantine, with $B=f\cdot N$ the number of Byzantine agents.
Each honest agent \(i\) maintains a scalar proposal \(\smash{v_i^{(t)} \in [0,50]}\), initialized at \(t=0\) by sampling i.i.d.\ from a fixed distribution over \([0,50]\) (uniform in our implementation), whereas Byzantine agents do not have an initial proposal and may choose arbitrary values in each round.
The goal of the protocol is to find agreement among honest agents through a structured exchange.

\paragraph{Protocol.}
Our simulator (A2A-Sim) shown in \cref{fig:bcg_overview} delivers structured messages over a synchronous network.
We present the full pseudo-code in \cref{app:a2a-sim} and describe the prompting strategy in \cref{app:prompts}.

\paragraph{Termination and outcomes.}
The simulator declares termination when at least \(2/3\) of all agents vote to stop; otherwise, the simulator times out at \(t = T_{\max}\) and we declare no consensus.
At termination, the game yields: (i)~\emph{valid consensus} (all honest agents hold identical values drawn from the initial honest proposals); (ii)~\emph{invalid consensus} (termination without validity); or (iii)~\emph{no consensus} (timeout). This is a \emph{no-stake} agreement game: the final agreement value carries no external reward for any of the agents, and evaluation focuses on agreement properties rather than optimizing the value.

\paragraph{Threat model.}
For simplicity, we consider a restricted Byzantine model.
Byzantine agents may propose arbitrary values and justifications and adapt to the history, but they cannot equivocate (send different messages to different recipients), forge identities, or drop or suppress messages.
In each round, a Byzantine agent must send the same message to all recipients. We demonstrate that even this straightforward threat model can have a substantial impact on consensus performance.

\paragraph{LLM agent implementation.}
Unless otherwise noted, all experiments use a common configuration.
Each agent uses the same prompt template specifying its role and inputs: local summary, current proposal, and private strategy from the previous round. The agent must return a JSON-formatted output containing a scalar proposal, a free-text justification, and a termination decision. We pass a compact textual summary of previous rounds rather than the full raw history to stay within the context window.
The summary includes, for the previous round, each agent's scalar proposal and a truncated justification, along with the focal agent's current proposal and private strategy.

\paragraph{Evaluation.}
For each configuration we run 25 independent simulations and record outcome rates (valid, invalid, premature stop, no consensus), rounds to termination, and consensus quality. We use \(T_{\max}=50\) for all experiments.
Outcome rates are reported with 95\% Wilson confidence intervals  over 25 runs \citep{wilson1927probable}. We test agents based on the Qwen3-8B/14B  model family \citep{yang2025qwen3}.

\section{Experiments and Results}

\subsection{Consensus without Byzantine agents}
We first study whether LLM-agent groups can reliably solve the scalar consensus game in benign conditions ($f=B=0$).
We run 600 simulations using A2A-Sim, varying agents $N \in \{4, 8, 16\}$ and evaluating Qwen3-8B and 14B under two prompt variants: one mentioning that Byzantine agents \emph{may exist} and one omitting any reference to Byzantine agents (see \cref{app:prompts} for more details).
Even without Byzantine agents, only 41.6\% of runs terminate in valid consensus.
Qwen3-14B substantially outperforms Qwen3-8B (67.4\% vs.\ 15.8\%), yet both exhibit high timeout rates.
Removing any mention of Byzantine agents from the prompt improves Qwen3-14B performance from 59.1\% to 75.4\% valid consensus and halves convergence time, suggesting that mentioning possible Byzantine agents harms liveness even when none are present.
Increasing group size further degrades performance: valid consensus drops from 46.6\% at $N{=}4$ to 33.3\% at $N{=}16$.
\cref{fig:combined_rounds_outcomes} summarizes outcomes across conditions.

\subsection{Consensus with Byzantine agents}
\begin{figure}[t]
\centering
\includegraphics[width=1\linewidth]{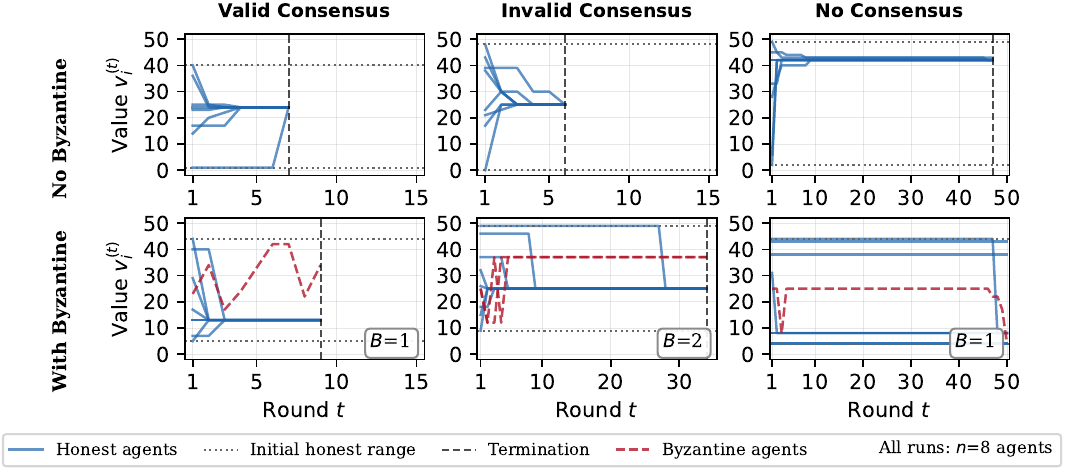}
\caption{Representative proposal trajectories for Qwen3-14B with eight honest agents. Top row: prompts explicitly state that no Byzantine agents exist. Bottom row: prompts include Byzantine agents and warn that Byzantine peers may exist. Each panel shows honest agents’ scalar proposals over rounds, with horizontal lines marking the initial honest range and a vertical line marking termination.}
\label{fig:trajectory}
\end{figure}

Next, we fix eight honest Qwen3-14B agents and add $B \in \{1,2,3,4\}$ Byzantine agents to assess robustness, corresponding to Byzantine fractions from $f = 1/9$ up to $f = 1/3$. \cref{fig:q2_byzantine_agents_comparison} highlights that invalid consensus remains rare across all settings, so Byzantine agents primarily harm liveness by preventing agreement rather than steering outcomes to corrupted values. \cref{fig:trajectory} further illustrates how threat-aware prompts and Byzantine agents shape proposal trajectories and can stall convergence even when validity is preserved.

\section{Discussion and Conclusion}

Even in benign, no-stake settings without Byzantine agents, LLM-agent groups frequently fail to reach \emph{valid} consensus within the round limit, and performance declines as group size increases. Under adversarial conditions, the likelihood of valid consensus decreases further, with failures primarily resulting from no-consensus outcomes, even within our limited threat model. These findings indicate that current LLM agents are not yet reliable social decision-makers: agreement, which is essential for cooperation, delegation, and safety-critical coordination, remains fragile in our controlled, no-stake testbed. Our study is limited by testing only a single Byzantine strategy and two model sizes from one family, and future work should investigate diverse adversarial behaviors, heterogeneous agent populations, and larger-scale deployments. We hope this work will inspire further research to address in greater depth the fundamental question of whether AI agents can reliably achieve agreement.

\bibliography{references}

@article{pease1980reaching,
  author  = {Pease, M. C. and Shostak, R. E. and Lamport, L.},
  title   = {{Reaching Agreement in the Presence of Faults}},
  journal = {{Journal of the ACM}},
  volume  = {27},
  number  = {2},
  pages   = {228--234},
  year    = {1980}
}

@article{lamport1982byzantine,
  author  = {Lamport, L. and Shostak, R. and Pease, M.},
  title   = {{The Byzantine Generals Problem}},
  journal = {{ACM Transactions on Programming Languages and Systems}},
  volume  = {4},
  number  = {3},
  pages   = {382--401},
  year    = {1982}
}

@inproceedings{li2023camel,
  author    = {Li, G. and Hammoud, H. A. A. K. and Itani, H. and Khizbullin, D. and Ghanem, B.},
  title     = {{CAMEL: Communicative Agents for ``Mind'' Exploration of Large Language Model Society}},
  booktitle = {{Advances in Neural Information Processing Systems (NeurIPS)}},
  year      = {2023}
}

@inproceedings{liu2024agentbench,
  author    = {Liu, X. and Yu, H. and Zhang, H. and Xu, Y. and Lei, X. and Lai, H. and others},
  title     = {{AgentBench: Evaluating {LLM}s as Agents}},
  booktitle = {{International Conference on Learning Representations (ICLR)}},
  year      = {2024}
}

@inproceedings{wu2024autogen,
  author    = {Wu, Q. and Bansal, G. and Zhang, J. and Wu, Y. and Li, B. and Zhu, E. and others},
  title     = {{AutoGen: Enabling Next-Gen {LLM} Applications via Multi-Agent Conversations}},
  booktitle = {{Conference on Language Modeling (COLM)}},
  year      = {2024}
}

@inproceedings{chen2024blockagents,
  author    = {Chen, B. and Li, G. and Lin, X. and Wang, Z. and Li, J.},
  title     = {{BlockAgents: Towards Byzantine-Robust {LLM}-Based Multi-Agent Coordination via Blockchain}},
  booktitle = {{ACM Turing Award Celebration Conference}},
  year      = {2024}
}

@misc{chen2023multiagent,
  author        = {Chen, H. and Ji, W. and Xu, L. and Zhao, S.},
  title         = {{Multi-Agent Consensus Seeking via Large Language Models}},
  year          = {2023},
  eprint        = {2310.20151},
  archivePrefix = {arXiv},
  note          = {{arXiv:2310.20151}}
}

@inproceedings{cemri2025why,
  author    = {Cemri, M. and Pan, M. Z. and Yang, S. and Agrawal, L. A. and Chopra, B. and Tiwari, R. and others},
  title     = {{Why Do Multi-Agent {LLM} Systems Fail?}},
  booktitle = {{Advances in Neural Information Processing Systems (NeurIPS)}},
  year      = {2025}
}

@misc{zheng2025rethinking,
  author        = {Zheng, L. and Chen, J. and Yin, Q. and Zhang, J. and Zeng, X. and Tian, Y.},
  title         = {{Rethinking the Reliability of Multi-agent System: A Perspective from Byzantine Fault Tolerance}},
  year          = {2025},
  eprint        = {2511.10400},
  archivePrefix = {arXiv},
  note          = {{arXiv:2511.10400}}
}

@misc{jo2025byzantinerobust,
  author        = {Jo, Y. and Park, C.},
  title         = {{Byzantine-Robust Decentralized Coordination of {LLM} Agents}},
  year          = {2025},
  eprint        = {2507.14928},
  archivePrefix = {arXiv},
  note          = {{arXiv:2507.14928}}
}

@misc{grotschla2025agentsnet,
  author        = {Gr\"otschla, F. and M\"uller, L. and T\"onshoff, J. and Galkin, M. and Perozzi, B.},
  title         = {{AgentsNet: Coordination and Collaborative Reasoning in Multi-Agent {LLM}s}},
  year          = {2025},
  eprint        = {2507.08616},
  archivePrefix = {arXiv},
  note          = {{arXiv:2507.08616}}
}

@misc{wolf2025this,
  author        = {Wolf, L. and Yoon, S. and Bogunovic, I.},
  title         = {{This Is Your Doge, If It Please You: Exploring Deception and Robustness in Mixture of {LLM}s}},
  year          = {2025},
  eprint        = {2503.05856},
  archivePrefix = {arXiv},
  note          = {{arXiv:2503.05856}}
}

@misc{yang2025qwen3,
  author        = {Yang, A. and Li, A. and Yang, B. and Zhang, B. and Hui, B. and Zheng, B. and others},
  title         = {{Qwen3 Technical Report}},
  year          = {2025},
  eprint        = {2505.09388},
  archivePrefix = {arXiv},
  note          = {{arXiv:2505.09388}}
}

@inproceedings{kwon2023efficient,
  author    = {Kwon, W. and Li, Z. and Zhuang, S. and Sheng, Y. and Zheng, L. and Yu, C. H. and others},
  title     = {{Efficient Memory Management for Large Language Model Serving with PagedAttention}},
  booktitle = {{Proceedings of the Symposium on Operating Systems Principles (SOSP)}},
  year      = {2023}
}

@article{wilson1927probable,
  author  = {Wilson, E. B.},
  title   = {{Probable Inference, the Law of Succession, and Statistical Inference}},
  journal = {{Journal of the American Statistical Association}},
  volume  = {22},
  number  = {158},
  pages   = {209--212},
  year    = {1927}
}
\bibliographystyle{iclr2026_conference}

\newpage
\appendix
\section{Appendix}

\subsection{Reproducibility} 
\label{app:reproducibility}
To ensure reproducibility, we provide: (1) complete experimental setup details; (2) full description of the A2A-Sim protocol, agent prompts, and Byzantine strategies; (3) comprehensive metrics and evaluation procedures; and (4) aggregated experimental results in the code repository:

\begin{center}
\href{https://github.com/ETH-DISCO/can-ai-agents-agree}{\texttt{https://github.com/ETH-DISCO/can-ai-agents-agree}}.
\end{center}

All experiments were conducted using vLLM~\citep{kwon2023efficient} for batched inference and guided-decoding. Models were obtained from Hugging Face and served with FP16/BF16 precision, 8{,}192 tokens context window. 

\subsection{A2A-Sim algorithm}
\label{app:a2a-sim}
The pseudo-code of A2A-Sim is shown in \cref{alg:a2asim-round}.

\begin{algorithm}[ht]
\caption{One round of A2A-Sim for agent $i$}
\label{alg:a2asim-round}
\begin{algorithmic}[1]
\Require local history $h_i^{(t)}$, current proposal $v_i^{(t)}$, role (honest or Byzantine)
\State Receive a textual summary of round $t{-}1$, local state, and messages from other agents.
\State Query the LLM policy with $(h_i^{(t)}, v_i^{(t)})$ to obtain a new proposal $\hat v_i^{(t+1)}$ and justification $r_i^{(t+1)}$.
\If{$i$ is Byzantine}
  \State arbitrarily modify $(\hat v_i^{(t+1)}, r_i^{(t+1)})$.
\EndIf
\State Broadcast $m_{i\to j}^{(t+1)} = (\hat v_{i\to j}^{(t+1)}, r_{i\to j}^{(t+1)})$ to every agent $j$.
\State Receive messages $M_i^{(t+1)} = \{m_{j\to i}^{(t+1)} : j=1,\dots,N\}$.
\State Update $h_i^{(t+1)} \gets h_i^{(t)} \cup M_i^{(t+1)}$ and set $v_i^{(t+1)} \gets \hat v_i^{(t+1)}$.
\State Query the LLM to emit a termination vote $\text{stop}_i^{(t+1)} \in \{\texttt{vote}, \texttt{continue}\}$
\end{algorithmic}
\end{algorithm}

\subsection{Prompt listings}
\label{app:prompts}

In this section, we present the exact prompts used to instantiate the behavior of honest and Byzantine agents during the experiments. Figure~\ref{prompt:byz_system_prompt} defines the global system prompt for Byzantine agents, specifying their adversarial role. Figure~\ref{prompt:byz_round_prompt} shows the prompt used at each communication round for Byzantine agents, while Figure~\ref{prompt:byz_round_vote_prompt} details how they are instructed to cast votes during rounds. Figure~\ref{prompt:byz_sys_vote_prompt} provides the system-level voting prompt governing their final decisions. For honest agents, Figure~\ref{prompt:hon_round_prompt} presents the round interaction prompt, and Figure~\ref{prompt:hon_round_vote_prompt} specifies how honest agents express their votes. The system prompts controlling honest agent behavior are shown in Figure~\ref{prompt:hon_sys_prompt_may_awareness} and Figure~\ref{prompt:hon_sys_prompt_none_exist}, which differ based on whether agents are informed about the possible presence of adversaries. Finally, Figure~\ref{prompt:hon_sys_vote_may_exist} and Figure~\ref{prompt:hon_sys_vote_none_exist} define the corresponding system-level voting prompts under these two assumptions.

\begin{figure}[h]
    \centering
    \displayprompt{assets/byz_system_prompt.tex}
    \caption{Byzantine system prompt defining the global adversarial behavior and strategy for Byzantine agents. Variable fields include: agent ID, group size, and allowed adversarial actions.}
    \label{prompt:byz_system_prompt}
\end{figure}

\begin{figure}[h]
    \centering
    \displayprompt{assets/byz_round_prompt.tex}
    \caption{Byzantine round prompt specifying the instructions for a single communication round. Variable fields: current round number, previous messages, and observed agent states.}
    \label{prompt:byz_round_prompt}
\end{figure}

\begin{figure}[h]
    \centering
    \displayprompt{assets/byz_round_vote_prompt.tex}
    \caption{Byzantine round vote prompt detailing how a Byzantine agent selects a vote during a round. Variable fields: candidate values, round number, and internal adversarial strategy parameters.}
    \label{prompt:byz_round_vote_prompt}
\end{figure}

\begin{figure}[h]
    \centering
    \displayprompt{assets/byz_sys_vote_prompt.tex}
    \caption{Byzantine system-level voting prompt that governs the final decision-making of Byzantine agents. Variable fields: group consensus state, prior votes, and adversarial override options.}
    \label{prompt:byz_sys_vote_prompt}
\end{figure}

\begin{figure}[h]
    \centering
    \displayprompt{assets/hon_round_prompt.tex}
    \caption{Honest agent round prompt defining instructions for message generation in a single round. Variable fields: agent ID, round number, prior messages, and group size.}
    \label{prompt:hon_round_prompt}
\end{figure}

\begin{figure}[h]
    \centering
    \displayprompt{assets/hon_round_vote_prompt.tex}
    \caption{Honest agent round vote prompt specifying how an honest agent selects a value to vote on. Variable fields: candidate values, current round, and message history.}
    \label{prompt:hon_round_vote_prompt}
\end{figure}

\begin{figure}[h]
    \centering
    \displayprompt{assets/hon_sys_prompt_may_awareness.tex}
    \caption{Honest system prompt (may-aware) controlling agent behavior when adversaries may be present. Variable fields: agent ID, group size, potential adversary count, and prior votes.}
    \label{prompt:hon_sys_prompt_may_awareness}
\end{figure}

\begin{figure}[h]
    \centering
    \displayprompt{assets/hon_sys_prompt_none_exist.tex}
    \caption{Honest system prompt (no-adversary) defining instructions when no adversaries are expected. Variable fields: agent ID, group size, and prior vote history.}
    \label{prompt:hon_sys_prompt_none_exist}
\end{figure}

\begin{figure}[h]
    \centering
    \displayprompt{assets/hon_sys_vote_may_exist.tex}
    \caption{Honest system vote prompt (may-exist) specifying final vote selection rules when adversaries may exist. Variable fields: candidate values, detected adversary influence, and previous round votes.}
    \label{prompt:hon_sys_vote_may_exist}
\end{figure}

\begin{figure}[h]
    \centering
    \displayprompt{assets/hon_sys_vote_none_exist.tex}
    \caption{Honest system vote prompt (no-adversary) specifying final vote selection when no adversaries are present. Variable fields: candidate values, round number, and prior messages.}
    \label{prompt:hon_sys_vote_none_exist}
\end{figure}

\end{document}